\documentclass[%
 reprint,
 amsmath,amssymb,
 aps,
]{revtex4-2}
\usepackage{changes}
\usepackage{graphicx}
\usepackage{dcolumn}
\usepackage{bm}

\begin{document}

\preprint{APS/123-QED}

\title{ Vortex Interference Enables Optimal 3D Interferometric Nanoscopy}

\author{Wei Wang $^{1,3}$}
\author{Zengxin Huang $^{1,3}$}
\author{Yilin Wang $^{1}$}
\author{Hangfeng Li$^{1}$}
\author{Pakorn Kanchanawong$^{1,2}$}%
 \email{biekp@nus.edu.sg}
\affiliation{%
 $^1$Mechanobiology Institute, Singapore 117411, Republic of Singapore\\
 $^2$Department of Biomedical Engineering, National University of Singapore, Singapore 117583, Republic of Singapore\\
  $^3$These authors contributed equally to this work
}%

\begin{abstract}
Super-resolution imaging methods that combine interferometric axial (z) analysis with single-molecule localization microscopy (iSMLM) have achieved ultra-high 3D precision and contributed to the elucidation of important biological ultrastructures.  However, their dependence on imaging multiple phase-shifted output channels necessitates complex instrumentation and operation.  To solve this problem, we develop an interferometric super-resolution microscope capable of optimal direct axial nanoscopy, termed VILM (Vortex Interference Localization Microscopy).  Using a pair of vortex phase plates with opposite orientation for each dual-opposed objective lens, the detection point-spread functions (PSFs) adopt a bilobed profile whose rotation encodes the axial position.  Thus, direct 3D single-molecule coordinate determination can be achieved with a single output image.  By reducing the number of output channels to as few as one and utilizing a simple 50:50 beamsplitter, the imaging system is significantly streamlined, while the optimal iSMLM imaging performance is retained, with axial precision 2 times better than the lateral. The capability of VILM is demonstrated by resolving the architecture of microtubules and probing the organization of tyrosine-phosphorylated signalling proteins in integrin-based cell adhesions.

\end{abstract}

\maketitle

Intensive development in super-resolution fluorescence microscopy in recent years have enabled their applications to facilitate a broad range of biological discoveries\cite{betzigscience,zhuangstorm,kanchanawong2010nanoscale,xu2013actin}. Among super-resolution microscopy techniques widely utilized in bioimaging, single-molecule localization microscopy (SMLM) depends on the stochastic spatiotemporal control of the density of the emitting fluorophores and their subsequent localization\cite{zhuangstorm,balzarotti2017nanometer,hell1994breaking,westphal2005nanoscale,hao2021three}. However, while SMLM can readily approach molecular-scale lateral (xy) precision especially with recent improvements in data analysis and labelling strategies, achieving comparable axial (z) precision is significantly more challenging. Commonly used approaches to realize 3D super-resolution in SMLM such as astigmatic imaging offers optical simplicity\cite{zhuang3d,pavani2009dh} but at the cost of at least 2-3 times poorer z-precision compared to the lateral (xy) precision. Greater extent of axial precision enhancement generally requires interferometric techniques, among which the configurations based on 4Pi geometry have been established to provide ultra-high 3D precision. Interferometric SMLM (iSMLM) techniques such as iPALM\cite{shtengel2009}, 4Pi-SMS\cite{aquino2011two}, W-4PiSMSN\cite{huang2016ultra,zhang2020nanoscale}, and 4Pi-STORM\cite{bates2022optimal}, provide a major gain in axial precision, averaging up to twice the lateral precision over the focus range, and are theoretically reaching the intrinsic quantum limit of spatial localization precision\cite{backlund2018fundamental}. The ultra-high precision of iSMLM methods have been instrumental in elucidating major biological insights\cite{kanchanawong2010nanoscale,huang2016ultra,bertocchi2017nanoscale,case2015molecular}.  However, in significant part due to the highly complex optical instrumentation, calibration, and operation\cite{shtengel2009,aquino2011two,huang2016ultra}, these techniques have remained confined to just a handful of laboratories world-wide, even after more than a decade since their emergence. To enable broader access to the research community thus requires the simplification of its operation while retaining its high optical performance. 

\begin{figure}[t]	
	\centering\includegraphics[width=8cm]{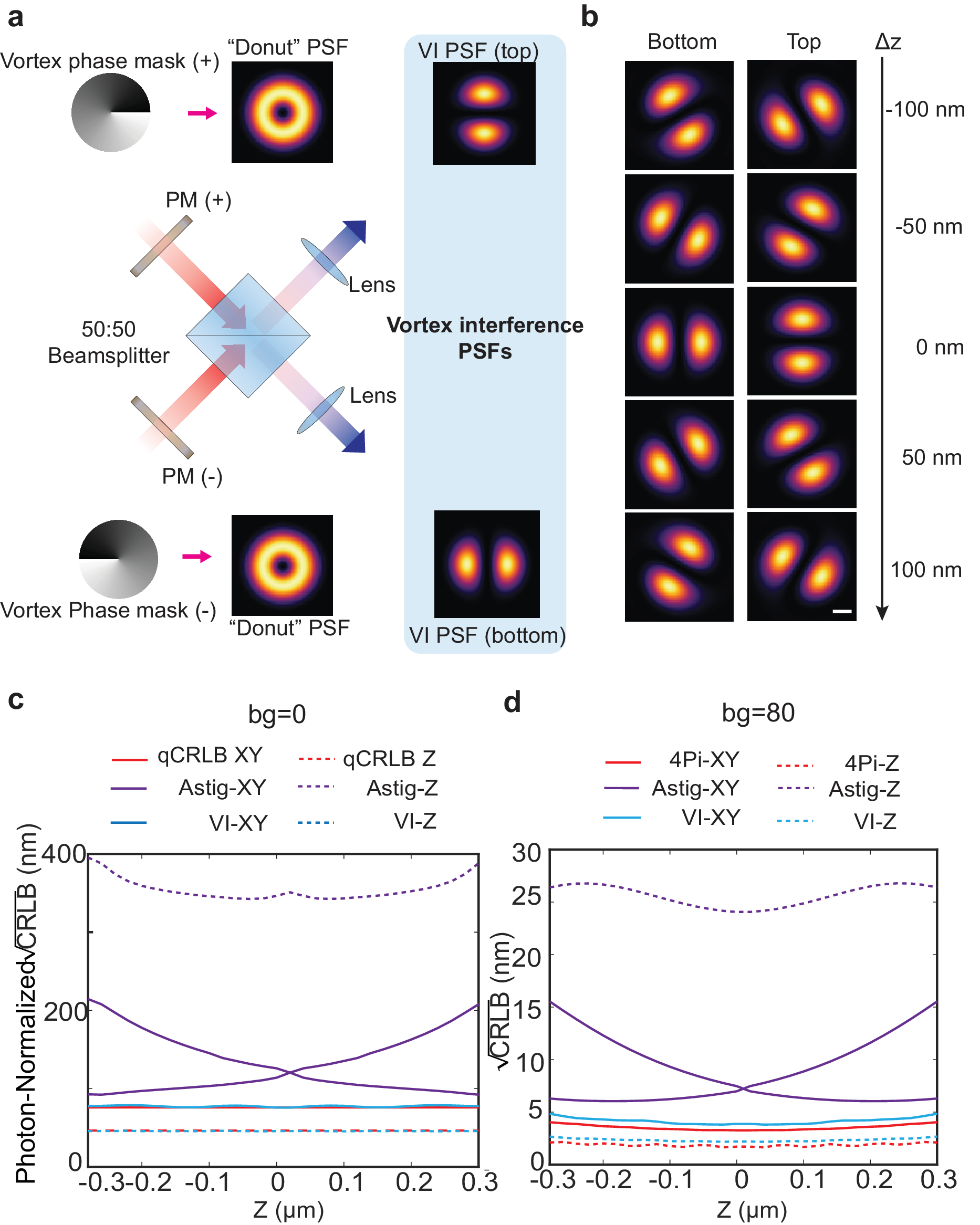}
	\caption{Principle of vortex interference (a) Schematic diagram depicting how light beams shaped by vortex phase masks (PM) with opposite vortex number (+/-) generate interference signals after the 50:50 beamsplitter (BS), giving rise to distinct bilobed profile (VI PSF).  (b) Axial (z)-dependent azimuthal rotation of de-magnified VI PSFs at indicated z-position. (670 nm emission, NA: 1.33, mag: 60, pixel size: $7.2\mu$m). (c) Comparison of analytical quantum bound precisions (qCRLB) and theoretical 3D precisions of VI and astigmatism PSFs. The precision is photon-normalized. (d) Comparison of precisions of 4Pi PSF, astigmatism PSFs and VI PSFs under a typical background photon value. Photon: 2000, total background photon per pixel (bg): 80. For astigmatism-based SMLM, both signal and background photon number is half as only one objective lens used). Scale bar: 200 nm, b.}
	\label{fig:1}
\end{figure}

 In current iSMLM methods, the 4Pi geometry encodes high-precision z-position information via self-interference of fluorescence emission. The axial position of a single-molecule fluorophore is then retrieved primarily via multi-phase interferometry\cite{shtengel2009,aquino2011two,huang2016ultra}. However, at least three detection channels are needed to solve the inverse problem\cite{shtengel2009}. As a result, complex optical configuration is required to realize three or four detection channels. For example, in iPALM, a custom-designed 3-phase beam splitter is used to achieve three detection channels\cite{shtengel2009}, while in 4Pi-SMS or 4Pi-STORM customized phase modulation devices are employed to realize four polarization-dependent channels\cite{aquino2011two,huang2016ultra}.
 In practice, achieving and maintaining optimal interference across multiple channels simultaneously is challenging. By reducing the number of detection channels to one or two and using a simple beamsplitter, the system complexity can be significantly reduced, enhancing the stability of the interference contrast. However, the problem of optimizing interferometric nanoscopy using a single beamsplitter has been relatively underexplored.
 
 To solve this problem, here we developed Vortex Interference Localization Microscopy (VILM), which is demonstrated to enable optimal direct 3D interferometric nanoscopy. By harnessing vortex beam interference, 3D spatial information of a single-molecule fluorophore can be retrieved using just one output image. By placement of a vortex phase mask in each 4Pi optical arm with opposite chirality, the self-interfered PSFs generated by a simple 50:50 beamsplitter adopt a bilobed profile with its orientation encoding the z-coordinate. 3D information is thus readily extracted using one channel. Our numerical analysis established that PSFs generated by VILM perform at the optimal limit for a single beamsplitter interferometric system. Biological imaging performance of VILM is validated by resolving the architecture of microtubules as well as signalling components of integrin-based cell adhesions.

\textbf{Theoretical foundation}  In a 4Pi configuration utilizing two opposed objective lenses, interference detection captures nearly the full $4\pi$ solid angle, resulting in a detection PSF with an axial fringe structure having three or four times finer feature.  In order to determine the axial position, multiple intensity measurements are required (see also Supplementary Notes) \cite{supVILM}. This can be achieved by various multi-phase projection methods which enable multiple output channels to be simultaneously imaged. Such multi-phase projection optics impose additional complexity and cost, however. Towards simplifying iSMLM, we realized that a significant design simplification can be achieved using vortex beam interference generated by commercially available vortex phase masks. 

In super-resolution microscopy, the vortex phase mask  $T=exp(±i\phi)$ is best known as the optical component that generates the “donut” beam used in STED or MINFLUX microscopy\cite{westphal2005nanoscale,balzarotti2017nanometer}. When 2 donuts with opposite handedness interfere, this gives rise to a pair of counter-rotating bilobes (see also Supplementary Notes I, Supplementary Movie 1)\cite{supVILM}.  This specific type of interference is referred to as "vortex interference" \cite{wisniewski2014mechanical,guo2016measuring}. As illustrated in Fig.1(a), vortex interference can be readily incorporated into a 4Pi geometry. Under a scalar approximation of the optical system, with a simple 50:50 beamsplitter, the detection PSFs for such configuration can be approximated as: 
\begin{eqnarray}
I_i = A_0 \dfrac{\rho_p^2}{\zeta^2}e^{-4ln2 \dfrac{\rho_p^2}{\zeta^2}}(1-cos(\psi_i+2\beta kz-2\varphi_p))
\end{eqnarray}
where $(\rho_p, \varphi_p)$  correspond to the polar coordinates in the image plane, and $\psi_i$  denotes the phase shifts between the light collected by the two channels. Here, $\beta$ and $\zeta$ are constants defined by the imaging system, while $A_0$  denotes the amplitude.  Accordingly, the PSF appears as a bilobed profile with two maxima and two minima, which azimuthally rotate as a function of the axial position (z) (Fig. 1(b)). The range for a complete rotation corresponds to approximately $\lambda/(2n_s)$   where $n_s$ is the refractive index of the specimen. Importantly, the axial position of the fluorophore can be directly calculated from the rotation angle of a bilobed PSF.  While two channels are generated as outputs, their information content is largely redundant, and duplex channel summation primarily improved precision due to higher photon counts but is otherwise dispensable (Supplementary Fig. S1(d)) \cite{supVILM}.  This configuration, here termed VILM (Vortex Interference Localization Microscopy), thus enables interferometric monoplex 3D coordinate read-out with simplified optical construction.

\begin{figure}[t]	
	\centering\includegraphics[width=8.5cm]{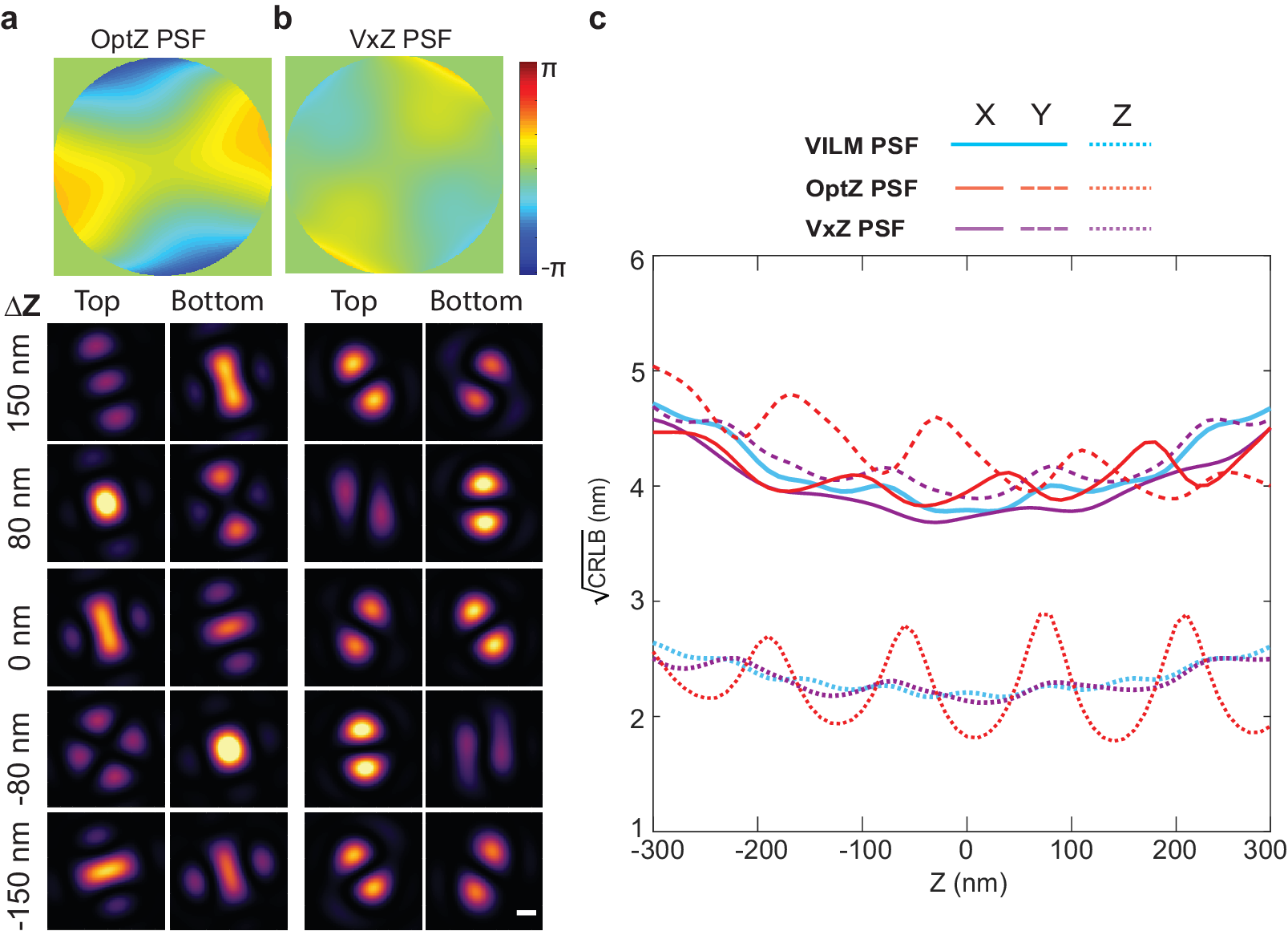}
	\caption{Optimal PSFs for 3D interferometric SMLM with single beamsplitter. (a-b) Optimal Zernike (OptZ) pupil function calculated using pure Zernike modes (a: top) and Optimal Vortex+Zernike pupil functions (VxZ) calculated using both Vortex and Zernike modes (b: top), and corresponding de-magnified PSFs at different z-position (bottom). (c). Comparison of theoretical 3D precisions for VILM PSFs, OptZ PSFs and VxZ PSFs. Photon: 2000, and total background photon per pixel: 80. Scale bars (a-b): 200 nm.}
	\label{fig:1}
\end{figure}

To evaluate the performance of VILM, we calculate Cramer-Rao Lower Bound (CRLB) theoretical localization precision in comparison to other super-resolution microscopy modalities as displayed in Fig. 1(c-d) \cite{ober2004localization,middendorff2008isotropic}.
As demonstrated in Ref. \cite{backlund2018fundamental}, the interferometric detection of single molecules achieves the fundamental quantum bound 3D precision when considering only Poisson noise\cite{helstrom1969quantum,tsang2016quantum}. As shown in Fig. 1(c), with zero background photon, the photon-normalized axial precision achieved by the VILM PSF is identical to the analytical quantum bound axial precision, while the lateral precision is nearly identical as well. In contrast, the precision achieved by astigmatism-based PSF\cite{zhuang3d}, a non-interference-based 3D method, is significantly worse than that of VILM PSF, with the axial precision being approximately 2 times worse than the lateral precision. Indeed, the theoretical axial precision of VILM is expected to be 2 times superior to the lateral precision, similar to other iSMLM methods. Importantly, unlike the traditional 4Pi system, which requires multiple phase shifts \cite{shtengel2009,aquino2011two,huang2016ultra}, VILM PSF achieves quantum-bound axial precision with just one beam splitter.

When the total background photon per pixel (bg) is increased to 80, a typical value observed during cell imaging, both the axial and lateral precisions of VILM are slightly worsen by approximately 1.2 times when compared to 4Pi PSF (Fig. 1(d)). However, VILM still provides significant localization precision enhancement over astigmatic imaging. For example, the theoretical axial and lateral precisions for monoplex VILM is 3 and 7 nm, respectively, whereas for astigmatic imaging, the axial and lateral precision are 20 nm and 7 nm, being significantly worse axially as described earlier (Supplementary Fig. S1(c-d) ) \cite{supVILM}.  Duplex VILM provides further precision enhancement with axial and lateral precision of 2 and 4 nm, respectively. Thus,VILM using only one beam splitter configuration achieves nearly the same 3D precision as conventional iSMLM under typical imaging conditions.

\begin{figure}[htbp]	
	\centering\includegraphics[width=8cm]{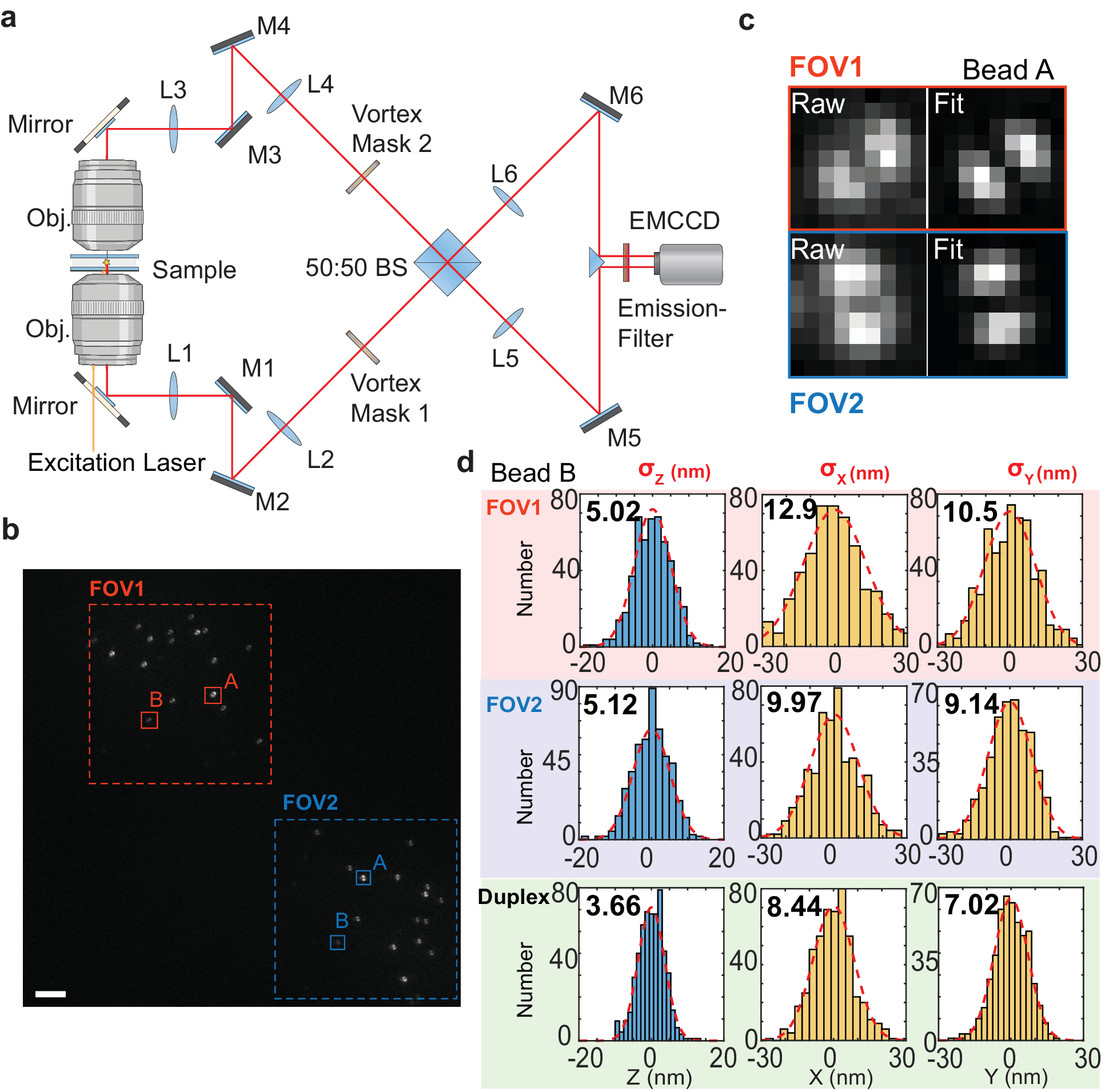}
	\caption{Vortex Interference Localization Microscopy (VILM) (a) Optical layout of VILM system.  (b) Raw image from duplex detection with top and bottom imaging areas (dashed) indicated by FOV1 and FOV2. (c) Example of experimental VILM PSFs. (Left) Raw images of a fluorescent fiducial A from solid boxed regions in FOV1 and FOV2 in (b). (Right) Fitted image. (d) Greater z-axis precision was observed for both monoplex and duplex detection. Histograms of fitted 3D positions of a typical fluorescent fiducial B comparing duplex (bottom row) and monoplex analysis of top (FOV1) and bottom (FOV2) channels. Gaussian fit (red dashed lines) and $\sigma$ of the fits are indicated. Scale bar: 5 $\mu$m.  }
	\label{fig:1}
\end{figure}

\begin{figure*}[htbp]	
	\centering\includegraphics[width=16cm]{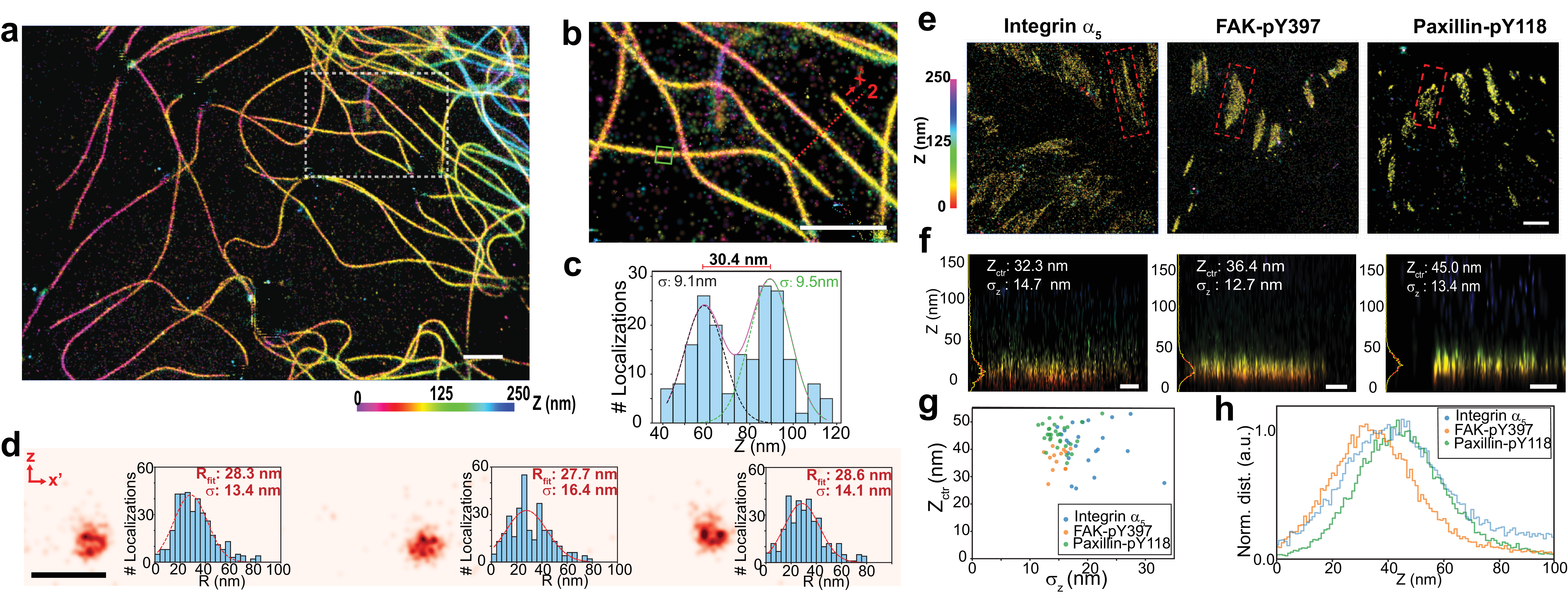}
	\caption{3D Nanoscopy of cellular structures by VILM (a) Reconstructed 3D super-resolution image of microtubule in a COS7 cell. Hue scale indicates z-dimension. (b) Zoomed-in view of boxed region in (a) with specified areas highlighted in (c, d) indicated by green box (1) and red dashed line (2), respectively. (c) Histogram of axial positions from boxed area in (b). Two distinct peaks are discerned with Gaussian-fitted axial precision indicated. (d) Reconstructed isometric transverse view of three microtubules along the axial and cross-section line indicated in (b). Corresponding histograms of radial distribution of localizations relative to the estimated center of microtubules and Gaussian fit parameters are shown. (e) Reconstructed 3D super-resolution images of integrin-based focal adhesion components in U2OS cells. (f) Transverse cross-section along longitudinal axes of red-dashed boxed regions in (e), along with histogram of z-positions (1-nm bin), and Gaussian fits (red) and fit parameters (centre and standard deviation: $z_{ctr}$ and $\sigma_z$).  Z-coordinate is referenced against a plane defined by substrate-embedded fiducials. (g) Distribution of $z_{ctr}$ and $\sigma_z$ for different focal adhesion components. Each datapoint corresponds to an adhesion region analyzed as depicted in (f). (h) Normalized cumulative histograms of z-position for adhesion regions in (g). Scale bars:  2 $\mu$m (a,b,e), 100 nm (d) and 500 nm (f).}
	\label{fig:3}
\end{figure*}

\textbf{Vortex interference enables optimal direct axial nanoscopy} We further evaluate the performance of VILM following a design strategy previously developed to identify the optimal 3D SMLM pupil function that generates a PSF with the best Fisher information properties in the presence of background photons\cite{shechtman2014optimal}.  We first calculated the optimal Zernike modes for realizing iSMLM with a single beamsplitter, with the emphasis on optimizing the axial precision as much as possible, while not seeking to extend the axial range.  The optimization problem is to minimize the weighted mean precision defined as follows:
\begin{equation}
\min \limits_{T}\{ \frac{1}{3N_z}\sum_i\sqrt{CRLB_i(z)}+ \alpha \max \limits_{z \in Z}(\sqrt{CRLB_z(z)})\} 
\label{eq:eq7}
\end{equation}
with the last term introduced to mitigate fluctuations in the axial precision and emphasize attaining the best possible axial precision. Here, $N_z$ is the number of sampling points in the axial direction and the coefficient $\alpha$ varies between 0 and 1. In addition to the optimal pupil function ($T$) based on solely linear combination of Zernike modes, we also calculated the optimal pupil function based on the linear combination of the vortex phase and Zernike modes. The two optimized pupil functions and corresponding z-depth-dependent PSFs are presented in Fig. 2(a-b), denoted OptZ PSF and VxZ PSF, respectively (see also Supplementary Notes V) \cite{supVILM}. Theoretical precision for these optimized PSFs as well as for the PSF of VILM are displayed in Fig. 2(c) and Supplementary Movies 1-3. Analogous to VILM PSFs whose intensity variations are transferred to the azimuthal direction, OptZ PSFs exhibit intensity variation along the lateral direction as a function of z-position. In the case of VxZ PSFs, the axial rotation is also maintained, along with additional intensity variation between the two channels. Compared to OptZ PSFs, we found that VILM PSFs offer comparable precision overall, with slightly better performance in axial (z) dimension. The axial precision of VILM PSF also vary more smoothly as a function of z-position compared to the greater undulation in OptZ PSFs. Comparison between VILM PSFs and VxZ PSFs reveals a similar trend. Notably, VxZ PSFs shows greater undulation in axial precision and anisotropy in lateral precision, primarily due to the intensity imbalance between the two channels in VxZ PSFs. Furthermore, while VILM enable single-channel detection due to the uniform intensity variation over the axial range, in practice both OptZ and VxZ would require dual-channel detection due to the intensity oscillation between the upper and lower output channels (see also Supplementary Notes II, Supplementary Movies 1-3) \cite{supVILM}. Taken together, we conclude that VILM theoretically attains the optimal super-resolution performance under the simplest single beamsplitter condition.

\textbf{Optical implementation and imaging performance validation.} To validate our design, we next constructed a VILM imaging system according to an optical configuration illustrated in Fig. 3(a) (see also Supplementary Fig. S4 and Supplementary Table S1) \cite{supVILM}. The system is based on a 4Pi geometry with dual opposed objective lenses aligned co-axially above and below the specimen plane. The specimen is mounted on a piezoelectric sample holder.  Subsequent to each objective lens, an assembly of relay lenses and mirrors are used to map the pupil plane to a vortex phase mask. Opposite orientation of the vortex phase mask for the upper and lower arms then generates vortex beam interference as described above. A simple 50:50 non-polarizing beam splitter (BS, Fig. 3(a)) is positioned to project the interfered emission signals toward the detection EMCCD camera, with monoplex or duplex detection modes selectable by the placement of either a mirror or a prism in front of the camera. Detailed steps in optical instrumentation, alignment, calibration, and operation of VILM are described in Supplementary Methods section \cite{supVILM}. We first image nanoparticle fluorescent fiducials to evaluate the imaging performance of VILM, using the duplex imaging mode to assess whether designed PSFs behave as expected.  As shown in Fig. 3(b-c), over both field-of-views (FOVs) each fiducial can be seen to exhibit bilobed shape consistent with the theoretical prediction.  The bilobed PSF orientations within each FOV are largely uniform since the substrate-affixed fiducials are co-planar in z. The PSFs corresponding to each fiducial are perpendicularly oriented between FOV1 and FOV2 as expected.  As the specimen is axially translated by the piezo sample holder, the rotation of bilobed PSFs are apparent (Supplementary Movies 4-5) with a complete rotation over half of the emission wavelength, as shown in Supplementary Fig. S6 \cite{supVILM}. Altogether, these results demonstrate that vortex interference effectively convert z-coordinate information into azimuthal rotation of the bilobed PSFs, hence encoding 3D coordinate information in a single-image.

To determine the 3D coordinate from the bilobed PSF, we made use of cubic spline fitting \cite{babcock2017analyzing,li2018real,bates2022optimal} based on theoretical PSFs calculated using the system parameters. As shown in Fig. 3(c), the bilobed PSF can be fitted well with theoretical model. The histograms of the 3D coordinates of a representative fiducial, with typical photon number of 1100-1200 in each channel, analyzed by either monoplex or duplex detection are presented in Fig. 3(d). The monoplex axial precision ($\sigma_z$) for FOV1 and FOV2 are 5.02 nm and 5.12 nm, respectively. In duplex mode, the axial precision is enhanced (3.66 nm) as the number of photons is roughly doubled. The monoplex lateral precision of FOV1 and  FOV2 are (12.99 nm, 10.55 nm) and (9.97 nm, 9.14 nm) respectively. Combining two channels together, the duplex lateral precision is enhanced accordingly (8.44 nm, 7.02 nm). For each channel, both the monoplex and duplex axial precision are about 2 times better than the lateral precision as defined by $\sqrt{\sigma_x \sigma_y}$.  Altogether, our results demonstrated that both Monoplex and Duplex VILM enable significant axial precision enhancement exceeding the lateral precision. Our analysis over an extended imaging depth (Supplementary Fig. S6 in \cite{supVILM}) shows that the axial precision is largely uniform within the central period.

 To demonstrate its performance in realizing direct axial nanoscopy, monoplex VILM is used to image cellular structures. The microtubule cytoskeletons are commonly used as a resolution benchmark in super-resolution microscopy. When labelled by immunostaining using unlabelled primary antibody and fluorophore-labelled secondary antibody, microtubule have been imaged as a hollow structure\cite{aquino2011two,huang2016ultra}.
 Compared to iPALM which required a long calibration and alignment time, the simplified procedure of VILM enabled a cell to be imaged within $3-5$ min after identification.  As shown in Fig. 4(a-b), VILM imaging of a fixed COS7 cell with microtubules immunostained using AlexaFluor 647$-$conjugated secondary antibody reveals clear 3D profiles of the filament networks. From the histogram of the axial-coordinates within the green box in Fig. 4(b), two distinct peaks, separated by 30.4 nm, can be observed  with $\sigma$ of 9.1 nm and 9.5 nm, respectively, as shown in Fig. 4(c). The hollow structure of microtubules can be discerned when the transverse cross-section reconstructed images along the red dashed line in Fig. 4(b) are viewed, with examples shown in Fig. 4(d).

Earlier, whether endogenous proteins or actual signalling events such as protein tyrosine phosphorylation are located in signalling components of integrin-based cell adhesions has not been directly determined.  Hence, we employed VILM to probe the nanoscale localization of key indicators of integrin-dependent signalling such the phosphorylation of Tyrosine 397 in FAK (FAK-pY397) and Tyrosine 118 in Paxillin (Paxillin-pY118). As shown in Fig.4(e-h), monoplex VILM imaging reveals that Integrin $\alpha$5$\beta$1, FAK-pY397, and Paxillin-pY118 form thin plaques in close proximity to the substrate plane and one another. Reconstructed images of the longitudinal projection reveal axially well-defined distribution, with a standard deviation of $\simeq13$ nm (Fig. 4(f-g) that reflect the compound axial precision and effective size of the primary/secondary antibody complexes. The analysis of the z-position histograms of adhesion regions revealed that z-center positions are distributed in the 35-45 nm range, consistent with earlier studies\cite{kanchanawong2010nanoscale}. Taken together, these results established the utility of VILM in resolving nanoscale organization in cells with high axial resolution.

In this study, we demonstrate how vortex interference can greatly simplify the optical implementation and operation of 3D interferometric nanoscopy. By encoding axial coordinate information into a rotating bilobed point spread function (PSF), we show that a single output image is sufficient for determining 3D single-molecule coordinates. Our theoretical analysis reveals that the VILM PSF achieves optimal performance with a single beamsplitter configuration. Finally, imaging of biological specimens showcases its effectiveness in performing direct interferometric 3D nanoscopy.

We thank Harald Hess, Gleb Shtengel (Howard Hughes Medical Institute) and Mikael Backlund (University of Illinois Urbana-Champaign) for helpful discussion. We acknowledge funding support by Quantum Engineering Programme (QEP-P7), Ministry of Education Academic Research Fund Tier 3 (MOE-T3-2020-0001), Mechanobiology Institute intramural support, and Singapore National Research Foundation Mid-Sized Grant (NRF-MSG-2023-0001) to P.K. H.L. is supported by MBI Graduate Scholarship. 

\providecommand{\singleletter}[1]{#1}%

\nocite{*}
\bibliography{bibvilm}

\end{document}